\documentstyle[12pt]{article}

\begin{document}

\baselineskip=18.6pt plus 0.2pt minus 0.1pt

\makeatletter
\@addtoreset{equation}{section}
\renewcommand{\theequation}{\thesection.\arabic{equation}}

\title{
\vspace{-2cm}
\rightline{\mbox
{\normalsize {Lab/UFR-HEP/0008}}}
\vspace{2cm}
\bf Non Trivial Extension of the (1+2)-Poincar\'e Algebra and Conformal Invariance on the Boundary of $\bf AdS_3$ }

\author{I.Benkaddour,A.El Rhalami and E.H.Saidi        \\
\small{Lab/UFR.High Energy Physics. Physics Department, Faculty of Science. }\\
\small{Av. Ibn Battota, B.P.1014, Rabat, Morocco}}

\maketitle
\hoffset=-1cm\textwidth=15cm
\vspace*{0.5cm}
\begin{abstract}
 Using recent results on strings on $ AdS_3\times N^d$, where N is a
  d-dimensional compact manifold, we re-examine the derivation of the
   non trivial extension of the (1+2)dimensional-Poincar\'e algebra obtained
    by Rausch de Traubenberg and Slupinsky, refs[1]and [29] . We show by
     explicit computation that this new extension is a special kind of
      fractional supersymmetric algebra which may be derived from the
       deformation of the conformal structure living on the boundary of
        $AdS_3$. The two $ so(1,2)$ Lorentz modules of spin $\pm{1\over k}$
         used in building of the generalisation of the $(1+2)$ Poincar\'e
          algebra are re-interpreted in our analysis as highest weight
           representations of the left and right Virasoro symmetries on the
            boundary of $ AdS_3$.  We also complete known results on
              2d-fractional supersymmetry by using spectral flow of
               affine Kac-Moody  and superconformal symmetries. Finally we
                make preliminary comments on
              the trick of introducing Fth-roots of g-modules to generalise
               the $so(1,2)$ result to higher rank Lie algebras g.\\

\bigskip


\end{abstract}
\newpage
\section{Introduction}
Recently a non trivial generalisation of the $(1+2)$ dimensional Poincar\'e algebra going beyond the standard supersymmetric extension has been obtained in [1]. In addition to the usual Poincar\'e generators, this extension refered herebelow to as the Rausch de Traubenberg-Slupinski algebra (RdTS algebra for short), involves two kinds of conserved charges $Q^{\pm}_{s}$ transforming as $so(1,2)$ Verma modules of spin $s={\pm {1\over k}};{k\geq 2}$. This construction is interesting first because it goes beyond standard 2d-fractional supersymmetry based on considering k-th roots of the $so(2)$ vector and second because it gives a new algebraic structure which a priori is valid for higher rank Lie algebras g where $ so(2)$ and $so(1,2)$ appear just as two special examples. In two dimensions where conformal invariance is infinite we now know, by help of conformal field theory methods and techniques of complex analysis, how to deal with objects type k-th root of $so(2)$ vector. For higher space time dimensions however, computations are in general difficult to perform except for some special situations such as the problem we will study herebelow and where RdTS symmetry find applications in low dimensional physical systems. In (1+2) dimensions, representations of the RdTS extension of the $ so(1,2)$ algebra have quantum states carrying fractional values of the spin and are expected to play a particular role in the exploration of special features of field theoretical models of (1+2)dimensional systems with boundaries. The idea of considering 3d systems with boundaries is crucial. It is motivated by the fact that one can imagine that the RdTS $ so(1,2) $ extension may naturally be linked to a 2d boundary conformal field theory (BCFT) living on the boundary of the space time. From this view we expect that RdTS construction for $ so(1,2) $ may be related to known results on integrable deformations of 2d conformal invariance. Recall that representations theory of conformal invariance in two dimensions [2] predict naturally the existence of quantum field operators generating states with exotic spins englobing the $ so(1,2) $ RdTS ones. It is then an interesting task to check if there exists effectively any relation between the RdTS generalisation of Poincar\'e invariance in (1+2) dimensions and known results on integrable deformations of $2d$ CFT's[3,4]. We expect that this relation exists really and its determination may help in understanding the behaviour of physical bulk quantities near the boundary of (1+2) dimensional systems. To study this problem we shall mainly work with $ AdS_3$ as the $ (1+2)$ space time with boundary and use recent results on strings propagating on $ AdS_3\times N^d $, where $N^d$ is a d-dimensional compact manifold to
be specified later on. The analysis we will develop in this paper might also be adapted to study some features of fractional quantum Hall(FQH) effects [5,6]; in particular the understanding of the correspondance between the
bulk effective Chern Simons (CS) gauge theory of FQH droplets and the conformal field theory living on its boundary [6,7].\\The aim of this paper is to exhibit explicitly the link between the RdTS analysis and 2d BCFT using recent results on D branes physics on the (1+2) dimensional anti de Sitter space $AdS_{3}$ [8,9,10]. We first show that there exists indeed a connection between the RdTS algebra and deformations of 2d space time BCFT. Then we establish the rule of correspondance between the two $so(1,2)$ Verma modules, used in constructing the non trivial extension of the (1+2) Poincar\'e invariance, and primary Virasoro representations of the full conformal algebra on the boundary of $AdS_3$. We show moreover that the RdTS supersymmetry, although obtained using an unusual method, has in fact the same origin as standard fractional supersymmetry (FSS) [11,12,13], see also[14,15]. Both FSS and RdTS algebras are residual subsymmetries of conformal invariance.\\
The presentation of this paper is as follows: In section 2, we review the basic ideas of FSS and RdTS supersymmetry using the conformal field theoretical method for the first and the algebraic approach for the second. We give explicit calculations for the deformation of the $C={4\over 5}$ Potts model. In section 3, we review the main lines of RdTS analysis. We also introduce some useful tools for the study of the link between the RdTS modules and highest weight representations(HWR) of the Virasoro algebra. In section 4, we study the relation between RdTS supersymmetry and two dimensional conformal invariance. We show in particular that the two $so(1,2)$ modules considered in building  supersymmetry are just special HWRs of the conformal invariance on the boundary of $AdS_3$. In section 5, we use the spectral flow of $2 d$ $N=2$ and $N=4$ superconformal invariances to complete the study of section 2 by giving a new result on FSS. We also take the opportunity of using spectral flow of affine Kac-Moody symmetries to give comments on the k-th roots of the $su(n)$ fundamental representations used by RdTS in extending their result for $so(1,2)$ for $ su(n)$. In sections 6 and 7, we give our results and  conclusion.
\section{ RdTS supersymmetry.}

 RdTS fractional supersymmetry is a special generalisation of FSS living in two dimensions and considered in many occasions in the past in connection with integrable deformations of conformal invariance and representations of the universal envelopping $U_q{sl(2)}$ quantum ordinary and affine symmetries[11,12,16,17]. Like for FSS, highest weight representations of RdTS algebra carry fractional values of the spin and obey more a less quite similar FSS eqs. We will show throughout this study that, up to some details related to the number of dimensions of space time, RdTS fractional supersymmetry has indeed the same origin as FSS. Both FSS and RdTS invariance  describe residual symmetries left after integrable deformations of scale invariance in two dimensions. To better understand the algebraic structure of FSS and RdTS supersymmetry we first propose to describe briefly the main lines of 2d FSS one gets from integrable deformations of conformal invariance. Then we give the RdTS extension of the $(1+2)$ dimensional Poincar\'{e} invariance as derived in [1].
\subsection{ 2d FSS }
FSS extends the usual Bose-Fermi symmetry in two dimensions; it exchanges bosons and quasiparticles (parafermions) of fractional spin instead of fermions. In addition to the energy momentum translation operator vector $P_{\pm}$, FSS is generated by conserved charges $Q_{x}$ and $\bar Q_{x}$ carrying fractional values of the spin x ( $x={l \over k};\quad 1< l <k  \quad mod[1] ;\quad k \geq 2$). These charge operators are remanant constants of motion that survive after integrable deformations of conformal invariance. There are various FSS algebras depending on the conformal model one starts with. For the example of the $Z_{k}$ parafermionic invariance of Zamolodchikov and Fateev (ZF)[18 ],see also [19], a way to get FSS algebras is as follows. First start from the ZF conformal algebra generated by the energy momentum tensor $T(z)$ and the parafermionic currents $\Psi{_q}(z), q=1,\ldots k$:\\ 
\begin{eqnarray} 
T_{\Psi }(z_1)T_{\Psi }(z_2)&=&{c_{\Psi }/2}z_{12}^{-4}+2{z_{12}^{-2}}T(z_2)+
{z_{12}^{-1}}\partial {T(z_2)}+\ldots\nonumber\\
\Psi_{k}(z_{1})\Psi_{l}(z_{2})&=&C_{k,l}^{k+l}{z_{12}^{-2kl/N}}\{ \Psi_{k+l}(z_{2})+\ldots \} ,\quad (k+l)<N ,\nonumber\\
\Psi_{k}(z_{1})\Psi_{k}^{+}(z_{2})&=&C_{k,N-l}^{N+k-l}{z_{12}^{-2k(N-l)/N}}\{ \Psi_{k-l}(z_{2})+\ldots \} ,\\
\Psi_{k}(z_{1})\Psi_{k}^{+}(z_{2})&=&z_{12}^{-2k(N-k)/N}
[1_{id}+{2\Delta _{k}/c_{k}}{z_{12}^{2}}T_{\Psi }(z_2)+\ldots],\nonumber\\
T_{\Psi }(z_{1}) \Psi_{k}(z_{2})&=&{\Delta_{k} \over z_{12}^{2}}\Psi_{k}(z_{2})+{1
\over z_{12}}\partial_z {\Psi_{k}(z_{2})}+\ldots \nonumber,
\end{eqnarray}
where the parameters $c_\Psi $ and $C_{k,l}^{k+l}$ are the central charges and structure constants  of the parafermoinic algebra respectively.
The $\Psi _{q}(z)$'s and the $\bar {\Psi_{q}}(\bar{z})$ have the conformal weights $\Delta_{q}=q{(k-q)\over k}$.
Second solve the following operator eqs:\\

\begin{eqnarray} 
P_{-}=\oint dz T(z)\nonumber\\
P _+=\oint d{\bar{z}}{\bar{T}(\bar{z})},
\end{eqnarray}
where $ T(z)$ and $\bar{T}(\bar{z})$ are replaced by their expressions in terms of the $\Psi ^{\pm}(z)$'s and the $\bar{\Psi} ^{\pm }(\bar{z})$ eqs(1). To solve these eqs, one has to specify the ZF parafermionic primary representations since the mode expansions of the $\Psi_q$'s and the $\bar{\Psi }_q$'s depend on the weight of the ZF primary field operators $ \Phi _q^p$.\\

\begin{eqnarray}
\Psi _{k}(z_1)\Phi _p^q(z_2)&=&{\sum _{n\in Z}} {z_{12}^{n-{kp/N}-k}Q _{-n+{k(p+k)\over N}}^{k,p} \Phi_p^q(z_2)} \nonumber\\
\Psi _{k}^{+}(z_1)\Phi _p^q(z_2)&=&{\sum _{n\in Z}} z_{12}^{n+{kp/N}-k}Q _{-n- {k(p+k)\over N}}^{-k,p} \Phi_p^q(z_2),
\end{eqnarray}
where $Q _{-n+{k(p+k)\over N}}^{k,p}$ and $Q _{-n- {k(p+k)\over N}}^{-k,p}$ are the modes of $\Psi _k$ and $\Psi _{k}^+$ respectively defined by:\\
\begin{eqnarray} 
Q _{-n+{k(p+k)\over N}}^{k,p}\Phi_p^q(z_2)&=&\oint dz_{1} z_{12}^{n+{kp/N+k-1}}\Psi (z_1)\Phi_p^q(z_2),\nonumber\\
Q _{-n- {k(p+k)\over N}}^{-k,p}\Phi_p^q(z_2)&=&\oint dz_{1} z_{12}^{n- {kp/N +k-1}}\Psi(z_1)\Phi_p^q(z_2).
\end{eqnarray}
To illustrate how things work in practice let us consider an example. The method we will present herebelow applies to all $Z_{k}$ parafermionic models as well as others such as the Tye et al symmetries [20,21 ].

\subsection{ Deformation of $\bf {C={4\over 5}}$ Potts model }

To fix the ideas, we consider the $c=4/5$ critical Potts model described by the following $Z_3$ parafermionic invariance. This is the leading non trivial example having constants of motion carrying fractional values of the spin. The algebra governing the critical behaviour of this model is:
\begin{eqnarray}
\Psi^{\pm}(z_{1})\Psi^{\pm}(z_{2})&\approx&{-z_{12}^{-2/3}}\Psi^{\pm}(z_{2})\nonumber\\
\Psi^{+}(z_{1})\Psi^{-}(z_{2})&\approx&{z_{12}^{-4/3}}[1+5/3{z_{12}^{2}}T(z_2)]\nonumber\\
T(z_{1})\Psi^{\pm}(z_{2})&\approx&{2/3\over z_{12}^{2}}\Psi^{\pm}(z_{2})+{1
\over z_{12}}\partial_z {\Psi^{\pm}(z_{2})}\\
T(z_1)T(z_2)&=&2/5 z_{12}^{-4}+2{z_{12}^{-2}}T(z_2)+
{z_{12}^{-1}}\partial {T(z_2)}\nonumber.
\end{eqnarray}
Similar relations are valid for $\bar{\Psi} ^{\pm}(\bar{z})$'s. The ZF parafermionic currents ${\Psi^\pm}$ have a spin $2/3$ and satisfy $([{\Psi^\pm}(z)]^+={\Psi^{\mp}}(z))$.\\
The algebra (2.4) has three parafermionic highest weight representations (PHWR)$[\Phi _{q}^q]$;\\$q=0,1,2$ namely the identity family $I=[\Phi _{0}^0]$ of highest weight $h_0=0$ and two degenerate families $[\Phi _1^1]$ and $[\Phi _2^2]$ of weights $h_1=h_2={1\over 15}$. Each one of these PHWRs is reducible into three Virasoro HWRs: $(\Phi_q^p); p=q, p=q{\pm 2}\quad (mod6)$. These field operators which are rotated omongst others under the action of the parafermionic currents as shown herebelow:
\begin{eqnarray}
{\Psi^{\mp}{\times }\Phi_q^p}&=&\Phi_q^{p \pm 2}\nonumber\\
\Phi_q^{p\pm 6}&=&\Phi_q^p ,
\end{eqnarray}
obey Virasoro and ZF primary conditions:\\
\begin{eqnarray}
L_n\vert{ h}\rangle&=&0,\quad n>o\nonumber\\
Q _{-n\pm(p\pm 1)/3}^{\pm}\vert{h}\rangle&=&0,{n\pm(p\pm 1)/3}>0,
\end{eqnarray}
where the $ L_n$ Virasoro and the $Q _{-n\pm(p\pm 1)/3}^{\pm}$ ZF modes are given by:
\begin {eqnarray}
L_n\vert{\Phi _p^q}\rangle&=&\oint dz z^{n+1}T(z)\Phi_p^q(0)\vert {0} \rangle,\nonumber\\
Q _{-n\pm(p\pm 1)/3}^{\pm}\vert{\Phi_p^q}\rangle&=&\oint dz z^{n\pm p/3}\Psi^{\pm}(z){\Phi_p^q}(0)\vert {0} \rangle.
\end{eqnarray}
Note that the the mode expansion of the ZF currents depend on the representation field operator on which they act. This property is manifestly seen on the energies of the creation and annihilation operators $Q _{-n\pm(p\pm 1)/3}^{\pm}$ which depend on the quantum number p of the ZF primary field $\Phi _q^{p}(z)$:

\begin{equation}
\Psi^{\pm}(z_1)\Phi _q^p(z_2)=\sum z_{12}^{n-1\mp{p/3}}Q _{-n\pm(p\pm 1)/3}^{\pm} \Phi_q^p(z_2),
\end{equation} 
 The ZF primary field operators $\Phi _q^p(z)$ satisfy also braiding properties type:

\begin{equation}
z_{12}^\Delta{\Phi _1(z_1)}{\Phi _2(z_2)}=z_{21}^\Delta{\Phi _2(z_2)}{\Phi _1(z_1)}=\Phi(z),
\end{equation}
 where $\Delta={\Delta_1}+{\Delta_2}-{\Delta_3},\Delta_i; i=1,2,3,$ are the conformal weights of the $\Phi_i$ field operators.\\
The second step in the derivation of FSS is to solve the operator eq(2.2) expressing the $2d$ energy momentum vector $P_{\pm}$ in terms of the ZF modes $Q _{-n\pm(p\pm 1)/3}^{\pm}$:

\begin{equation}
\begin{array}{lcr}
P _-=\oint dz {3\over 5}z^{-{2/3}}(\Psi^{+}(z)\Psi^{-}(0))\\
P _+=\oint d{\bar{z}}{{3\over 5}z^{-{2/3}}(\bar{\Psi}^{+}(\bar{z})\bar{\Psi}^{-}(0))}.
\end{array}
\end{equation}
where we replaced $ T(z)$ and $\bar{T}(\bar{z})$  in terms of the $\Psi^{\pm}(z)$'s and the $\bar{\Psi}^{\pm}(\bar z)$ as given by eqs(4 ). The solution of eqs(9) involves three pairs of doublets of the charge operators $(Q_{-{1\over 3}}^{\pm }, \bar{Q} _{1\over3}^{\pm } )$, $(  Q_{-{2\over3}}^{\pm }, \bar{Q} _{2\over3}^{\pm })$ and $({Q_{0}}^{\pm }, \bar{Q} _{0}^{\pm })$. Using the primary highest weight conditions (2.7), one can check by explicit computation that the $ Q , \bar Q, P_{-}$ and $P_+$ charge operators generate the following algebra:
\begin{eqnarray}
{\bf P}&=&Q_{-{1/3}}^{+}Q_{0}^{+}Q_{-{2/3}}^{+}\Pi_{0}+
Q_{-{2/3}}^{+}Q_{-{1/3}}^{+}Q_{0}^{+}\Pi_{1}+Q_{0}^{+}
Q_{-{2/3}}^{+}Q_{-{1/3}}^{+}\Pi_{-1}\nonumber\\
\lbrack {P_{\pm},Q_{- x}}\rbrack &=&0;\quad x=0,1/3,2/3 \nonumber \\
{\bf \bar{P}}&=&{{\bar{Q}}_{-{1/3}}}^{+}{{\bar{Q}}_{0}}^{+}
{{\bar{Q}}_{-{2/3}}}^{+}{\bar{\Pi}}_{0}+{{\bar{Q}}_{-{2/3}}}^{+}
{{\bar{Q}}_{-{1/3}}}^{+}{{\bar{Q}}_{0}}^{+}{\bar{\Pi} }_{1}+{{\bar{Q}}_{0}}^{+}{{\bar{Q}}_{-{2/3}}}^{+}{{\bar{Q}}_{-{1/3}}}^{+}{\bar{\Pi}}_{-1}\quad\\
\lbrack {P_{\pm},{\bar{Q}}_{+ x}}\rbrack &=&0\nonumber.
\end{eqnarray}
In these  eqs the $\Pi_{q}$'s and ${\bar{\Pi}}_{q} $'s are projector operators on the q-th ZF primary state $[{{\Phi _{q}}^q}{\times} {{\bar{\Phi}} _{q}}^q]$. The algebra (2.12) may also be obtained by analysing the energy spectrum of the mode operators $Q _{-n\pm{(p\pm 1)}/3}^{\pm}$ and $\bar{Q }_{-n\pm(p\pm 1)/3}^{\pm}$, n integer. The $Q_{-n\pm(p \pm 1)/3}^{\pm }$'s and $\bar {Q} _{-n\pm(p\pm 1)/3}^{\pm}$'s,  which depend on the $p$ charge, act only on the conformal representation $\vert{\Phi _p^q}\rangle$. This property may be interpreted to mean that expect the $\vert{\Phi _{p}^q}\rangle$ family, the action of the $Q_{-n\pm(p \pm 1)/3}^{\pm }$'s kills all states $\vert{{\Phi _p}^r}\rangle$ with r different from q. For $q=0$ for example, the non vanishing actions of $Q_{-x}^{\pm}$ and $\bar{Q}_{-x}^{\pm}, x=0,1/3,2/3$  on the states $\vert {s,p}\rangle$ of spin s, $0 {\leq} s {\leq}1 $ and charge $p$ read as:

\begin{eqnarray}
Q_{-{2/3}}^{\pm}{\vert{0,0}\rangle}&=&\vert{2/3,0}\rangle\nonumber\\
Q_ 0^{+}\vert{2/3,+2}\rangle &=&\vert{2/3,-2}\rangle \nonumber\\
Q_ 0^{-}\vert{2/3,-2}\rangle &=&\vert{2/3,+2}\rangle\\
Q_ {-{1/3}}^{+}\vert{2/3,-2}\rangle &=&\vert{1,0}\rangle \nonumber\\
Q_ {-{1/3}}^{-}\vert{2/3,+2}\rangle &=&\vert{1,0}\rangle \nonumber\\
\end{eqnarray}
and similar eqs for the antiholomorphic sector. From these Eqs as well as the expansion(2.3-4) of the ZF currents, one sees that $Q_ {-{1/3}}^{\pm}$ and $ Q_ {0}^{\pm}$ cannot act directly on the state $\vert{0,0}\rangle$. Similarly $Q_ {-{2/3}}^{\pm}$ cannot operate directly on $\vert{2/3,\pm 2}\rangle$. This result gives an explicit argument showing that FSS should be generated by more than one $Q $ and $\bar{Q}$ operators as it was naively used in earlier physical litterature on FSS. It shows moreover that not all the $Q_{-{x}}^{\pm}$'s are independent since we have:  

\begin{eqnarray}
Q_{-{1/3}}^{-}&=& Q_{-{1/3}}^{+}Q_{0}^{+}\nonumber \\
Q_{-{1/3}}^{+}&= &Q_{0}^{-}Q_{-{1/3}}^{-} \nonumber\\
Q_{-{2/3}}^{-}&=& Q_{0}^{+}Q_{-{2/3}}^{-}\\
 Q_{-{2/3}}^{+}&=& Q_{-{2/3}}^{-}Q_{0}^{-}\nonumber.
\end{eqnarray}
 Similar expressions may be written down for for the antiholomorphic sector. Putting back these relations into eqs(2.12), we find the following linearized algebra.

\begin{eqnarray}
 2P_{-1}&=&{\lbrace} {Q_{-{2/3}}^{+},
Q_{-{1/3}}^{-}} {\rbrace}+ {\lbrace}{Q_{-{1/3}}^{+},Q_{-{2/3}}^{-}}{\rbrace}\nonumber\\
0&=& {\lbrace} {Q_{-{1/3}}^{\pm},Q_{-{1/3}}^{\pm}}{\rbrace}=
{\lbrace}{Q_{-{2/3}}^{\pm},Q_{-{2/3}}^{\pm}}{\rbrace}.
\end{eqnarray}
We shall return to this linearised realisation of FSS in section5  when we discuss the spectral flow of $N=2$ and$ N=4$ superconformal invariance in two dimensions,wher a similar result will be obtained by using special choices of the parameter of the flow.
\section{ More on RdTS supersymmetry}

In this section we review briefly the derivation of the RdTS extension of the $(1+2)$ dimensional Poincar\'{e} invariance. We also initiate the study of a field realisation of RdTS supersymmetry which we develop further in the forthcoming section. In this regards we would like to note that as far as $SO(1,2)$ group is concerned, we will encounter in our analysis various kinds of  $SO(1,2)$ symmetries with different physical interpretations. In addition to the $SO(1,2)$ Lorentz invariance of the $(1+2)$ dimensional space time considered in [1], we will handle four $SO(1,2)$ invariances classified as:\\ (1) Two $SO(1,2)$'s given by the zero mode subgroup product  ${SO(1,2)}\times\bar{SO(1,2)}$ associated to ${so_{k}(1,2)}\times{\bar{so_{k}(1,2)}}$ affine Kac Moody invariance to be studied in section 4. This subsymmetry will be realised by using the usual $Sl(2,R)\sim SO(1,2)$ Wakimoto field theoretical representation[22].\\ (2) Two other $SO(1,2)$ subsymmetries associated to the non anomalous subalgebras of the left and right Virasoro  symmetries of some two dimensional BCFT of $AdS_3$ to be specified later on.
\subsection{RdTS extension of $\bf SO(1,2)$}
To start consider the Poincar\'{e} symmetry in $(1+2)$ dimensions generated by the space time translations $P_{\mu}$ and the Lorentz rotations $J_{\alpha}$ satisfying altogether the following closed commutation relations:
\begin {eqnarray}
\lbrack {J_{\alpha},J_{\beta}} \rbrack &=&i{\epsilon_{\alpha \beta \gamma}\eta^{\gamma \delta} P_ {\delta}}\nonumber\\
\lbrack {J_{\alpha},J_{\mu}}\rbrack &=&i{\epsilon_{\alpha \beta \gamma}\eta^{\gamma \delta}P_ {\delta}}\\
\lbrack {P_{\mu}, P_{\nu}}\rbrack&=&0\nonumber.
\end {eqnarray}
In these eqs, $\eta _{\alpha \beta}=diag (1, -1, -1)$ is the $(1+2)$ Minkowski metric and $\epsilon_{\alpha \beta \gamma}$ is the completly antisymmetric Levi-Civita tensor such that $\epsilon _{012}=1$. A convenient way to handle eqs(3.1) is to work with an equivalent formulation using the following Cartan basis of generators $P_{\mp}=P_1{\pm}i P_2$ and $J_{\mp}=J_1{\pm}i J_2$. In this basis eqs(3.1) read as:

\begin{eqnarray}
\lbrack {J_+, J_-}\rbrack &=& -2J_0 \nonumber\\     
\lbrack {J_0, J_{\pm}}\rbrack &=&  \pm {J_{\pm}}\nonumber\\ 
\lbrack {J_{\pm}, P_{\mp}}\rbrack &=& \pm{P_0}\\
\lbrack {J_+, P_+}\rbrack &=& \lbrack {J_-, P_-}\rbrack = 0 \nonumber\\
\lbrack {J_0, P_0}\rbrack &=& \lbrack {P_{\pm}, P_{\mp}}\rbrack = 0\nonumber.
\end{eqnarray}
The algebra (3.1-2) has two Casimir operators $P^{2}= {P_0}^2 - {1\over 2}(P_{+}P_{-} +P_{-}P_{+})$  and $P.J= P_{0}J_{0}- {1\over2}(P_{+}J_{-} +P_{-}J_{+})$. When acting on highest weight states of mass m and spin s, the eigenvalues of these operators are $m^2$ and $ms$ respectively. For a given s, one distinguishes two classes of irreducible representations: massive and massless representations. To build the $so(1,2)$ massive representations, it is convenient to go to the rest frame where the momentum vector $P_{\mu}$ is $(m,0,0)$ and the $SO(1,2)$ group reduces to its abelian $SO(2)$ little subgroup generated by $J_0$; $(J_{\pm}=0)$. In this case, massive irreducible representations are one dimensional and are parametrized by a real parameter. For the full $SO(1,2)$ group however, the representations are either finite dimensional for ${\vert {s} \vert}\in {\bf {Z}^{+}}/2$ or infinite dimensional for the remaning values of $s$.\\
 Given a primary state $\vert {s} \rangle $ of spin $s$, and using the abovementioned $SO(1,2)$ group theoretical properties, one may construct in general two representations HWR(I) and HWR(II) out of this state $\vert {s} \rangle$. The first representation HWR(I) is a highest weight representation given by:

\begin{eqnarray}
J^0{\vert {s}\rangle}&=& s{\vert {s}\rangle}\nonumber\\  
J_{-}{\vert {s}\rangle}&=& 0\nonumber\\
{\vert {s,n} \rangle}&=& \sqrt{{\Gamma (2s)}\over {\Gamma(2s+n)\Gamma (n+1)}}(J_{+})^n{\vert 
{s}\rangle}, n{\geq}1\\
J_{0}{\vert{s,n}\rangle}&=&(s+n){\vert{s,n}\rangle}\nonumber\\
J_{+}{\vert{s,n}\rangle}&=&\sqrt {(2s+n)(n+1)}{\vert{s,n+1}\rangle}\nonumber\\
J_{-}{\vert{s,n}\rangle}&=&\sqrt {(2s+n-1)n}{\vert{s,n-1}\rangle}\nonumber.
\end{eqnarray}
The second representation is a lowest weight representation which we refer to denote as HWR(II) is defined as:

\begin{eqnarray}
\bar{J _0}{\vert {\bar{s}} \rangle} &=& -s {\vert {\bar{s}}\rangle}\nonumber\\
    \bar{J_{+}}  {\vert {\bar{s}}\rangle}&=& 0\nonumber\\
{\vert\bar{s,n}\rangle}&=& (-)^{n}{\sqrt{{\Gamma(2s)}\over {\Gamma(2s+n)\Gamma(n+1)}}} 
(\bar{J_{-}})^{n} {\vert{\bar{s}}\rangle}\\
\bar{J _0}{\vert \bar{s,n} \rangle} &=& -(s+n) {\vert {\bar{s,n}}\rangle}\nonumber\\
\bar{J_+}{\vert\bar{s,n}\rangle}&=&{-\sqrt{(2s+n-1)n}}
\vert \bar{s,n+1}\rangle\nonumber.
\end{eqnarray}
Note in passing that in the second module we have supplemented the generators and the representations states with a bar index. This convention of notation will be justified later on. To fix the ideas, HWR(I) will be identified in section 7 with a left Virasoro Verma module and HWR(II) will be interpred as a right Virasoro one. Note moreover that both HWR(I) and HWR(II) representations have the same $so(1,2)$ Casimir $C_s$= $s(s-1), s<0$. For $s\in \bf Z^-/2$, these representations are finite dimensional and their dimension is $(2\vert {s} \vert+1)$. For generic real values of $s$, the dimension of the representations is however infinite. If one chooses a fractional value of $s$ say $s=-{{1}\over k}$; each of the two representations (3.3-4) splits a priori into two isomorphic representations respectively denoted as $D_{\pm 1/k}^+ and D_{\pm 1/k}^-$. This degeneracy is due to the redundancy in choosing the spin structure of $\sqrt {-{2/k}}$ which can be taken either as $+i\sqrt {-{2/k}}$  or $-i\sqrt {-{2/k}}$. These representations are not independent since they are related by conjugations; this why we shall use hereafter the choice of [1] by considering only 
 $D_{-{1/k}}^{+}$ and $D_{-{1/k}}^-$. In this case the two representation generators $J_{0,\pm}$ and $\bar{J}_{0,\pm}$ are related as:
\begin{equation}
\bar{J}_{0,\mp}=(J_{0,\pm})^*.
\end{equation}
Furthermore taking the tensor product of the primary states ${\vert {s} \rangle}$ and ${\vert {\bar{s}} \rangle}$ of the two $so(1,2)$ modules HWR(I) and HWR(II) and using eqs(3.3-4), it is straightforward to check that it behaves like a scalar under the full charge operator ${J_0}\times {1_{d}}+{1_{d}}\times {\bar{J}} _0$ which we denote simply as $J_0+{\bar{J}} _0$ :

\begin{equation}
(J_0+\bar{J}_0){\vert{s}\rangle}\otimes{\vert{\bar{s}}\rangle}= 0.
\end{equation}
Eq(3.6) is a familiar relation in the study of primary states of Virasoro algebra. This equation together with the mode operators $J_{-}^n $ and  $\bar{J}_{+}^{m}$ which act on ${\vert{s}\rangle}\otimes{\vert{\bar{s}}\rangle}$ as:

\begin{eqnarray}
(J_{-})^n{\vert{s}\rangle}\otimes{\vert{\bar{s}}\rangle}&=& 0,\quad n{\geq}1\nonumber\\
(\bar{J}_{+})^m {\vert{s}\rangle}\otimes{\vert{\bar{s}}\rangle}&=& 0,\quad m{\geq}1
\end{eqnarray}
define a highest weight state which looks like a Virasoro primary state of spin 2s and scale dimension $\Delta= 0 $. We will show later on when we study the primary field representation of the $2d$ BCFT of a string propagating on the $AdS_3$ background, that eq(3.6-7) correspond indeed to:

\begin{eqnarray}
({L _0}-{\bar{L} _0}){\Phi _{h,\bar{h}}}(0,0)
{\vert{0}\rangle}&=& (h-{\bar{h}}){\Phi _{h,\bar{h}}}(0,0)
{\vert{0}\rangle}\nonumber\\
({L _0}+{\bar{L} _0}){\Phi _{h,\bar{h}}}(0,0)
{\vert{0}\rangle}&=& (h+{\bar{h}}){\Phi _{h,\bar{h}}}(0,0)
{\vert{0}\rangle}\\
{L _n}{\Phi _{h,\bar{h}}}(0,0){\vert{0}\rangle}&=& 0\nonumber,\quad n{\geq}1\\
{\bar{L} _m}{\Phi _{h,\bar{h}}}(0,0){\vert{0}\rangle}&=& 0\nonumber,\quad m{\geq}1.
\end{eqnarray}
where $L _n$ and $\bar{L} _m$ are respectively the usual left and right Virasoro modes and $\phi_{h,\bar{h}}(z,\bar z)$ is a primary conformal field representation of conformal scale $h$+ $\bar{h}$ and conformal spin $h$- $\bar{h}$. This property, which gives the connection between RdTS supersymmetry and conformal invariance, will be explicited in details when we discuss HWRs of the conformal symmetry on the boundary of $AdS_3$. The primary $so(1,2)$ highest weight states $\vert{s}\rangle$ and $\vert{\bar{s}}\rangle$ eqs(3.3-4) are respectively in one to one correspondance with the left Virasoro primary state $\Phi _{h}(0){\vert{0}\rangle}
=\vert{h}\rangle$ and the right Virasoro primary one ${\Phi _{\bar{h}}}(0){\vert{0}\rangle}=\vert
{\bar{h}}\rangle$.\\ 
On the other hand, if we respectively associate to HWR(I) and HWR(II) the mode operators $Q_{s+n}^{+}= Q_{s+n}$  and $Q_{-s-n}^{-}=\bar{Q} _{s+n}$= and using $ SO(1,2)$ tensor product properties, one may build, under some assumptions, an extension $\bf S $ of the $ so(1,2)$ algebra going beyond the standard supersymmetric one. To do so, note first that the system { $J_0$, $J_+$ ,$J_-$ and  $Q_{s+n}$ obey the following commutation relations $(s=-1/k)$.

\begin{eqnarray}
\lbrack {J_0,Q_{s+n}}\rbrack &=&(s+n)Q_{s+n}\nonumber\\
\lbrack {J_+,Q_{s+n}}\rbrack &=&\sqrt {(2s+n)(n+1)}Q_{s+n+1}\\
\lbrack {J_-,Q_{s+n}}\rbrack &=&\sqrt {(2s+n-1)n}Q_{s+n-1}\nonumber.
\end{eqnarray}
Similarly we have for the antiholomorphic sector:

\begin{eqnarray}
\lbrack {{\bar{J}} _0,{\bar{Q}} _{s+n}}\rbrack &=&-(s+n){\bar{Q}} _{s+n}\nonumber\\
\lbrack {{\bar{J}} _+,{\bar{Q}} _{s+n}}\rbrack &=&-{\sqrt {(2s+n-1)n}}{{\bar{Q}} _{s+n-1}}\\
\lbrack {{\bar{J}} _{-},{\bar{Q}} _{s+n}}\rbrack &=&-{\sqrt {(2s+n)(n+1)}}{{\bar{Q}} _{s+n+1}}\nonumber.
\end{eqnarray}
To close these commutations relations with the $Q_s$'s through a k-th order product one should fullfil some constraints. \\(1) the generalized algebra $\bf S$we are looking for should be a generalisation of what is known in two dimensions, i.e a generalisation of FSS.\\(2) When the charge operator $Q_{s+n}$ goes arround an other, say $Q_{s+m}$, it picks a phase $\Phi ={2i{\pi} /k}$; i.e:

\begin{equation}
Q_{s+n}Q_{s+m}=e^{\pm 2i{\pi} s}Q_{s+m}Q_{s+n}+ \ldots;\quad  s= {-{1\over k}},
\end{equation}
where the dots refer for possible extra charge operators of total $J_0$ eigenvalue (2s+n+m). Eq (3.11) shows also that the algebra we are looking for has a ${\bf Z}_k$ graduation. Under this discrete symmetry, $Q_{s+n}$ carries a $+1 (mod k)$ charge while the $P_{0,\pm}$ energy momentum components have a zero charge mod k. \\(3) the generalised algebra $\bf S $ should split into a bosonic B part and an anyonic A and may be written as: ${\bf S} = {\oplus_{r=0}^{k-1}} A_r  = B {\oplus_{r=1}^{k-1}} A_r $ . Since $A_{n}A_m \subset  A_{(n+m) (modk)}$ one has:
\begin{eqnarray}
{\lbrace {A_r}{\ldots} {A_r} \rbrace}_{k}&{\subset}& {B}\nonumber\\
\lbrack {B,A}\rbrack &{\subset}& {B}\\
\lbrack {B,B}\rbrack &{\subset}& {B}\nonumber.
\end{eqnarray}
In these eqs, ${\lbrace {A_r}{\ldots} {A_r} \rbrace}_{k}$ means the complete symmetrisation of the k anyonic operators $A_r$ and is defined as: 

\begin{equation}
{\lbrace {A_{s_r}}{\ldots} {A_{s_r}\rbrace}_{k} ={1\over k!}{\sum _{\sigma {\in} {\Sigma}}}{(A_{s_{\sigma (1)}}{\ldots} {A_{s_{\sigma(k)}}}}},
\end{equation}
where the sum is carried over the k elements of the permutation group  $\lbrace {1, {\ldots},k}\rbrace$. \\(4) the algebra $\bf S$ should obey generalised Jacobi identities. In particular we should have:

\begin{equation}
adB{\lbrace {A_{s_1}}{\ldots} {A_{s_k}}\rbrace}=0,
\end{equation}
where B stands for the bosonic generators $J_{0,{\pm}}$ or $P_{0,{\pm}}$ of the Poincar\'{e} algebra. Using eq(3.12) to write ${\lbrace {A_r} {\ldots} {A_r}\rbrace}_k$ as $\alpha _{\mu}P^{\mu}+ \beta _{\mu}J^{\mu}$ where $\alpha$ and $\beta$ are real constants; then putting back into the above relation we find that ${\lbrace {A_r} {\ldots} {A_r}\rbrace}_k$ is proportional to $P_{\mu}$ only. In other words, $\beta _{\mu}$ should be equal to zero; a property which is easily seen by taking $B=P_{\mu}$ in eq (3.14). Put differently the symmetric product  of the $D_{s}^\pm$, denoted hereafter as $S^k[D_{s}^\pm]$, contains the space time vector representation $D_1$ of $so(1,2)$ and so the primitive charge operators $Q_{-{1/k}}$ and $\bar{Q}_{{1/k}}$ obey:

\begin{eqnarray}
\lbrack {J_0,(Q_{-{1/k}})^k} \rbrack&=&-(Q_{-{1/k}})^k{\sim} P_-\nonumber\\
\lbrack {J_-,(Q_{-{1/k}})^k} \rbrack&=&0.
\end{eqnarray}
Similarly we have:

\begin{eqnarray}
\lbrack {\bar{J}_0,({\bar{Q}}_{{1/k}})^k }\rbrack&=&({\bar{Q}} _{{1/k}})^k{\sim} P_+\nonumber\\
\lbrack {\bar{J} _+,({\bar{Q}} _{1/k})^k} \rbrack& =&0.
\end{eqnarray}
Moreover acting on $(Q _{-{1/k}})^k$ by $ad{J_{+}^n}$ and on $(\bar{Q} _{{1/k}})^k$ by $ad \bar{J}_{+}^n$, one obtains:
\begin{eqnarray}
{ad{J_+}} (Q _{-{1/k}})^k&{\sim}& P_0\nonumber\\  
 {ad{\bar{J}} _-} ({\bar{Q}} _{{1/k}})^k&{\sim}& P_0\\ 
 {{ad}^2 {J_+}} (Q _{-{1/k}})^k&{\sim}& {P_-}\nonumber\\
 {{ad}^2{\bar{J}} _-} ({\bar{Q}} _{{1/k}})^k&{\sim}& P_+\nonumber.
\end{eqnarray}
In summary, starting from $so(1,2)$ lorentz algebra (3.1-2) and the two Verma modules HWR(I) and HWR(II)(3.3-4 ), one may build the following new extended symmetry:
\begin{eqnarray}
{\{ Q_{-{1\over k}}^{\pm},\ldots ,Q_{-{1\over k}}^{\pm}\}}_{k}&=&P_{\mp}
=P_{1}{\pm}iP_{2}\nonumber\\
\{ Q_{-{1\over k}}^{\pm},\ldots ,Q_{-{1\over k}}^{\pm},Q_{1-{1\over k}}^{\pm}\}_{k} &=&\pm i\sqrt {2\over k}P_{0}-(k-1)\{ Q_{-{1\over k}}^{\pm},\ldots ,Q_{-{1\over k}}^{\pm},Q_{1-{1\over k}}^{\pm},Q_{1-{1\over k}}^{\pm}\}_{k}\nonumber\\
 &&\pm i\sqrt {k-2}\{ Q_{-{1\over k}}^{\pm},\ldots ,Q_{-{1\over k}}^{\pm},Q_{1-{1\over k}}^{\pm},Q_{2-{1\over k}}^{\pm}\}_{k}\nonumber\\
\lbrack J^{\pm},\lbrack J^{\pm},\lbrack J^{\pm},(Q_{-{1\over k}}^{\pm})^{k}\rbrack \rbrack \rbrack &=&0.
\end{eqnarray}
Eq(3.18) defines what we have been refering to as RdTS algebra. For more details on this algebraic structure, see [1,23].

\section{ Furthermore on RdTS supersymmetry}

Here we would like to answer the question rised in the introduction concerning the link between RdTS supersymmetry and two dimensional conformal invariance. We have anticipated on the nature of this link by saying that RdST supersymmetry is expected to arise from appropriate deformations of two dimensional CFT's on the boundary of $AdS_3$. The appearence of the $AdS_3$ space in this analysis is due to the fact that this geometry has many relevant features for our present study. We give hereafter two useful properties regarding the space time $SO(1,2)$ Lorentz group:\\ (1) In its euclidean representation, $AdS_3$ has an $SO(1,3)$ isometry group containing as subgroup the $SO(1,2)$ Lorentz symmetry of the (1+2) space time we are interested in. \\(2) The two dimensional $AdS_3$ boundary space may be realised as a two sphere on which may live boundary conformal field theories, which themselves have $so(1,2)$ projective subsymmetries that can be related to the above mentionned $so(1,2)$ Lorentz group.\\ Starting from these observations we want to show that the two so(1,2) modules HWR(I) and HWR(II), considered in the building of RdTS supersymmetry, are just special representations of the $AdS_3$ BCFT.
To proove this relation in a comprehensive manner, let us first review briefly some elements of $ AdS_3$ geometry. The $ AdS_3$ space is given by the hyperbolic coset manifold $Sl(2,C)/SU(2)$ which may be thought of as the three dimensional hypersurface ${H_3}^+$
\begin{equation}
-{X_0}^2+{X_1}^2+{X_2}^2+{X_3}^2= -l^2,
\end{equation}
embedded  in flat $ R^{1,3}$ with local coordinates { $X^0$, $X^1$, $X^2$, $X^3$ }. This hypersurface describes a space with a constant negative curvature ($-1\over {l^2}$). The parameter l is choosen to be quantized in terms of the $ {l_s }$ fundamental string lenght units; i.e, $l={l_s }\times k$ where k is an integer to be interpreted later on as the Kac Moody level of the $ {so_{k}(1,2)}$ affine symmetry. To study the field theory on the boundary of $AdS_3$, it is convenient to introduce the following set of local coordinates of $AdS_3$:
\begin{eqnarray}
\phi&=&{log(X_{0}+X_{3})}/ {l}\nonumber\\
{\gamma} &=&{{X_2+iX_0} \over {X_0+iX_3}} \\
{\bar{\gamma}} &=&{{X_2-iX_1}\over {X_{0}+iX_3}}\nonumber.
\end{eqnarray}
An equivalent description of the hypersurface is:
\begin{eqnarray}
{\gamma}&=&{r\over {\sqrt{l^2+r^2}}} e^{-\tau +i\theta }\nonumber\\
{\bar{\gamma}}&=&{r \over {\sqrt{l^2+r^2}}} e^{-\tau-i\theta}\nonumber\\
{\phi}&=&{\tau} +1/2log(1+r^2/l^2)\\
r&=&{le^{\phi}}\sqrt{\gamma \bar{\gamma}}\nonumber\\
\tau&=&\phi-1/2log(1+e^{2\phi}\gamma \bar{\gamma})\nonumber\\
{\theta} &=&{1\over {2i}} {log(\gamma /\bar{\gamma})}\nonumber,
\end{eqnarray}
where we have used the change of variables:
\begin{eqnarray}
X_0&=&X_0(r,\tau)={\sqrt{l^2+r^2}}cosh{\tau}\nonumber\\
X_3&=&X_3(r,\tau)={\sqrt{l^2+r^2}}sinh{\tau}\\
X_1&=&X_1(r,\theta)=rsin{\theta}\nonumber\\
X_2&=&X_2(r,\theta)=rcos{\theta}\nonumber.
\end{eqnarray}    
In the coordinates $( \phi, \gamma, \bar\gamma )$, the metric of $H_3^+$ reads as:

\begin{equation}
ds^2=k(d{\Phi}^2+e^{2{\Phi}}{d{\gamma}}{d{\bar{\gamma}}}).
\end{equation} 
 Note that in the $( \phi, \gamma, \bar\gamma )$ frame, the boundary of euclidean $AdS_3$ corresponds to take the field $\Phi $ to infinity. As shown on eq(4.3-4),this is a two sphere which is locally isomorphic to the  complex plane parametrized by $ (\gamma,\bar{\gamma})$ .\\
Quantum field theory on the $ AdS_3$ space is very special and has very remarkable features governed by the Maldacena correspondence in the zero slope limit of string theory[24 ]. On this space it has been shown that bulk correlations functions of quantum fields find natural interpretations in the conformal field theory on the boundary of $AdS_3$[9]. In algebraic language, this correspondance transforms world sheet symmetries of strings on $AdS_3$ into space time infinite dimensional invariances on the boundary of $AdS_3$. In what follows we shall review some useful properties of strings on $AdS_3$ and $\partial{AdS_3}$.
 
\subsection{ $\bf AdS_3$ / CFT correspondence.}

Strings propagating on the $AdS_3$ background are involved in the study of supersymmetric gauge theories with eight supercharges; in particular in the understanding of the Higgs and Coulomb branches near the moduli space singularity[25 ]. Strings on $ AdS_3$ have rich symmetries; some of them turn out to be related to the problem we are studing. These symmetries, which may be classified into WS symmetries and space time invariances, 
carry all relevent informations one needs to know about the string dynamics on $AdS_3$. In what follows we want to give some useful relations regarding these two classes of symmetries. To work out explicit field theoretical realisations of these symmetries, we start by recalling that in the presence of the Neveu-Schwarz $B_{\mu \nu}$ field with euclidean world sheet parameterized $(z,\bar{z})$, the dynamics of the bosonic string on $AdS_3$ is described by the following classical lagrangian:

\begin{equation}
 L=k[{\partial {\Phi}}{\bar{\partial}}{\Phi}+e^{2{\Phi}}{\partial {\gamma}}{\partial {\bar{\gamma}}}].
\end{equation}
 In this eq ${\partial}$ and $\bar{\partial}$ stand for derivatives with respect to z and $\bar{z}$ repectively. Introducing two auxiliary variables ${\beta}$ and ${\bar{\beta}}$, the above eq may be put into the following  convenient form:

\begin{equation}
 L^{'}=k^{2}({\partial {\Phi}}{\bar{\partial} {\Phi}}+{\beta {\bar{\partial}} \gamma}+{\bar{\beta}}{\partial {\bar{\gamma}}}-e^{-2{\Phi}}
{\beta {\bar{\beta}.}})
\end{equation}
The eqs of motion of the various fields one gets from eq(4.7) read as:

\begin{eqnarray}
{{\partial}{\bar{\partial}}}{\Phi} -2{\beta {\bar{\beta}}}e^{-2{\Phi}}&=&0\nonumber\\
{\bar{\partial}}\gamma-\beta e^{-2{\Phi}}&=&0\\
\partial{\bar{\gamma}}-{\bar{\beta}} e^{-2{\Phi}}&=&0\nonumber\\
\partial{\bar{\beta}}&=&{\bar{\partial}}\beta=0\nonumber.
\end{eqnarray}
 String dynamics on the boundary of $AdS_3$ is obtained from the previous bulk eqs by taking the limit $\Phi$ goes to infinity.This gives: 

\begin{eqnarray}
{{\partial}{\bar{\partial}}}\Phi &=&0\nonumber\\
{\bar{\partial}}\gamma&=&\partial {\bar{\gamma}}=0\\
\partial{\bar{\beta}}&=&\bar{\partial}\beta =0\nonumber.
\end{eqnarray}
The WS fields $\Phi, \gamma$ and $\bar{\gamma}$ which had general expressions in the bulk become now holomorphic on the boundary of $AdS_3$ and describe a  BCFT.
Note that consistency of quantum mechanics of the string propagating in space time requires that the target space should be $AdS_3{\times}N$, where N is a (3+n) dimensional compact manifold. To fix the ideas, N may be thought of as $S^3{\times}T^n$ with $n =20$ for the bosonic string and $n= 4$ for superstrings. We shall consider hereafter both of string and superstring cases.Given the big number of relations one may write down, we shall use however a strategy in which we give the strict necessary results. Thus our plan in what follows is: First, we describe some algebraic features of the WS invariance; then we make a pause to give a complement on FSS using spectral flow of $N=2$ and $N=4$ conformal invariance, after what we return to complete space time symmetries on the boundary of $AdS_3$and finally we give our results.

\subsection{WS Symmetries}

 World sheet invariances include affine Kac-Moody, Virasoro symmetries and their extensions. For a bosonic string propagating on $AdS_3{\times}S^3{\times}T^{20}$, we have the following:\\
{\bf A}- Three kinds of WS affine Kac-Moody invariances:\\
(a)A level $(k-2)$ ${sl(2)}\times{\bar{sl(2)}}$ invariance coming from the string propagation on $AdS_3$. This invariance is generated by the conserved currents $J_{sl(2)}^q$ and ${{\bar{J}}_{sl(2)}}^{q}; q=0,{\pm} 1$. In terms of the WS fields ${\Phi, \gamma, \bar{\gamma},\beta} $ and $\bar{\beta}$ of eq(?), the field theoretical realization of these currents is given by the Wakimoto representation:

\begin{eqnarray}
{J^{-}}(z)&=&{\beta}(z)\nonumber\\
J^{+}(z)&=&{\beta}{{\gamma}^{2}}+{\sqrt{2(k-2)}}{\gamma}{\partial {\Phi}} + k{\partial {\gamma}}\nonumber\\
J^{0}(z)&=&{\beta}{\gamma}+1/2{\sqrt{2(k-2)}}{\partial {\Phi}}\\
{\bar{J}}^{-}(\bar{z})&=&\bar{\beta}\nonumber\\
{\bar{J}^0}(\bar{z})&=&{\bar{\beta}}{\bar{\gamma}}+1/2{\sqrt{2(k-2)}}
{\partial {\Phi}}\nonumber \\
{{\bar{J}}^{+}}(\bar{z})&=&{\bar{\beta}}{\bar{\gamma}}^{2}+{\sqrt{2(k-2)}}{\bar{\gamma}}{\partial {\Phi}} + k{\partial {\bar{\gamma}}}\nonumber.
\end{eqnarray}
(b)A level $(k+2)$ invariance coming from the string propagation on $S^3$. The conserved currents are $J_{su(2)}^q$ and ${\bar{J}}_{su(2)}^q$. The WS field theoretical realization of these currents is given by the level $(k+2)$  WZW $su(2)$ model [26 ].\\
(c)A $u(1)^{20}{\times}{\bar{u}(1)}^{20}$ invariance coming from the torus $T^{20}$. This symmetry is generated by $20$ $ U(1)$ Kac Moody currents $J_{u(1)}^i;i=1,\ldots,20$.

{\bf B}-WS Virasoro symmetry

This symmetry, which splits into holomorphic and antiholomorphic sectors, is given by the Suggawara construction using quadratic Casimirs of the previous WS affine Kac Moody algebras. For the holomorphic sector, the WS Virasoro currents of a bosonic string on ${AdS_3}\times {S^3}\times {T^{20}}$ are:\\
(a) String on $ AdS_3 $:

\begin{equation}
T_{sl(2)}^{WS}={1\over (k-2)}[(J_{sl(2)}^0)^{2}-(J_{sl(2)}^{1})^{2}-(J_{sl(2)}^{2})^{2}].
\end{equation}

(b) String on $ S^3 $:

\begin{equation}
T_{su(2)}^{WS}={1\over (k+2)}[(J_{su(2)}^0)^{2}+(J_{su(2)}^{1})^{2}+(J_{su(2)}^{2})
^{2}]
\end{equation}

(c) String on $ T^{20} $:

\begin{equation}
T_{u(1)}^{WS}={\sum_{i=1}^{20}}{[J_{u(1)}^i]}^2.
\end{equation}

Similar quantities are also valid for the antiholomorphic sector of the conformal invariance. Note that the total WS energy momentum tensor $T_{tot}^{WS}$ is given by the sum of $T_{sl(2)}^{WS}, T_{su(2)}^{WS}$ and $T_{u(1)}^{WS}$ eqs(4.11-12-13).\\ In the case of a superstring propagating on $AdS_3{\times}S^3{\times}T^4$, the above conserved currents are slightly modified by the adjunction of extra terms due to contributions of WS fermions . If we denote by $\Psi_{sl(2)}^{A}, \Psi_{su(2)}^{a}$ and $\Psi_{u(1)}^{i}$, the $AdS_3, S^3$ and $T^4$ fermions, the WS theory has a N=1  superconformal theory generated by:
\begin{eqnarray}
T(z)&=&{1\over {k}}[(J_{sl(2)}^{A}J_{sl(2),A}-{\Psi_{sl(2)}^{A}}\partial \Psi_{sl(2),A})+ \nonumber\\&&(J_{su(2)}^{a}J_{su(2),a}-{\Psi _{su(2)}^{a}}\partial \Psi_{su(2),a})]+\nonumber\\&&1/2 {\sum _{i=1}^{4}}((J_{u(1)}^{i}J_{u(1)}^{i}-{\Psi_{u(1)}^{i}}\partial {\Psi_{u(1)}^{i}})\nonumber\\
G(z)&=&{2\over {k}}[{\Psi_{sl(2)}^{A}}J_{sl(2),A}-{i\over {3k}}\epsilon _{ABC}{\Psi_{sl(2)}^{A}}{\Psi_{sl(2)}^{B}}{\Psi_{sl(2)}^{C}}]+\nonumber\\&&{2\over {k}}[{\Psi_{su(2)}^{a}}J_{su(2),a}-{i\over {3k}}\epsilon _{abc}{\Psi_{su(2)}^{a}}{\Psi_{su(2)}^{b}}{\Psi_{su(2)}^{c}}]\nonumber\\&&+\sum _{i=1}^{4}{\Psi_{u(1)}^{i}}\partial J_{u(1)}^{i}.
\end{eqnarray}
Note that to get a space time supersymetric vaccum, one should enhance the previous N=1 superconformal WS invariance to a N=2 conformal symmetry [27]. This requires the existance of a conserved U(1) current in the world sheet theory under which G splits in two parts $G^+$ and $G^-$ with charges +1 and -1 respectively. Skiping the details and denoting by $G_r^\pm $ the modes of $G^{\pm}(z)$ N=2 fermions currents; the N=2 U(1) superconformal algebras read as:
\begin{eqnarray}
\lbrack { G_r^{-},G_s^{+}\rbrack}&=&2L_{r+s}-(r-s)J_{r+s}+(c/3)(r^2-1/4)\delta _{r+s,0}\nonumber\\
\lbrack L_n,L_m\rbrack &=&(n-m)L_{m+n}+{c\over {12}}m(m^2-1)\delta _{m+n,0}\nonumber\\
\lbrack L_n,G_r^{\pm} \rbrack &=&({n\over {2}} -r)G_{n+r}^{\pm}\\
\lbrack L_n,J_m \rbrack &=&-mJ_{m+n}\nonumber\\
\lbrack J_m,J_n \rbrack &=&{c\over {3}}m \delta _{m+n,0}\nonumber\\
\lbrack J_n,G_r^{\pm} \rbrack &=&\pm G_{n+r}^{\pm}\nonumber,
\end{eqnarray}
where the r and s modes take half odd integer values for the Neveu Schwarz(NS) sector and integer ones for the Ramond  (R)sector.
Before going ahead we would like to make a pause in order to give some relevant features of these algebras. this pause is motivated by the two following:
 First the $N=2$ NS and R conformal algebras have a spectral flow which we want to use in order to complete the study of section 2 on FSS by giving a new result. 
Second space time symmetry of superstring on $AdS_3\times S^3 \times T^4$ has a N=4 superconformal invariance which have a spectral flow of the same nature as for N=2 U(1) conformal invariance. Like for the FSS case, the spectral flow of $ N=2$ and $N=4$ conformal invariances may also be used to study RdTS supersymmetry.

\section{FSS and spectral flow}

In section 2, we have defined FSS as a hidden finite dimensional invariance which survives after integrable deformations of critical models such as the thermal deformation of $Z_N$ models; see eqs(2.11-12). There, we had exposed a method for deriving FSS algebras from parafermionic invariance. In the present section we want to complete the study of section 2 by giving a new way for obtaining FSS using topological field theory ideas [28]. This method is based on using an appropriate choice of the parameter $\eta$ of the spectral flow of $N= 2$ and $N= 4$ superconformal theories. We will also take the opportunity of analysing the spectral flow of $N=2$ and $N=4$ conformal symmetries to make a comment on the recent proposal of  [29] where a new construction of fractional supersymmetric algebras was derived by using infinite dimensional modules of Lie algebras.\\ To start recall that due to boundary conditions of fermions, the $2d N=2$ $( N=4)$ superconformal algebra has two sectors: Neveu Schwarz (NS) sector and Ramond (R) sector. These two sectors are not completely independent since they may be related by a continuous spectral flow as shown herebelow:
\begin{eqnarray}
U_ {\theta} L_n U_ {\theta}^{-1}&=& L_n+\theta J_n +{c/6} {\theta}^2 \delta_{n,0}\nonumber\\
U_ {\theta} J_n U_ {\theta}^{-1}&=& J_n +{c/3} {\theta} \delta_{n,0}\\
U_ {\theta} {G_r^+ }U_ {\theta}^{-1}&=&G_{r+\theta}^+\nonumber\\
U_ {\theta} {G_r^- }U_ {\theta}^{-1}&=&G_{r-\theta}^-\nonumber ,
\end{eqnarray}
           
for N=2 theories
\begin{eqnarray}
T_n^3(\eta )&=&T_n^3(0)-{{\eta kp}\over 2} \delta_{n,0}\nonumber\\
T_{n_{\pm}\eta}^\pm(\eta )&=&T_n^\pm (0)\nonumber\\
Q_{n+n/2}^1(\eta )&=&Q_n^1(0)\\
Q_{n-\eta/2}^2(\eta )&=&Q_n^2(0)\nonumber\\
L_n(\eta )&=&L_n(\eta )-\eta T_n^3(0)+\eta ^{2} ({kp\over 4 })  \delta_{n,0}\nonumber
\end{eqnarray}
               
for N=4 superconformal ones. The variable $\eta $ is the parameter of the spectral flow.                                                
Eqs (5.1-2) mean that $2d N=2$ $(N=4)$ superconformal algebras have then a continuous one parameter sector interpolating between NS and R algebras. This interpolating sector is generated by mode operators $G_{r{\pm} {\eta}}^{\pm}$ and ${\bar{G}}_{r{\pm} {\eta}}^{\pm}$ carrying shifted values of $L_0$ and the $U(1)$ charge operators. For a generic value of $\eta$, the commutation relations of the $N=2$  superconformal algebra in two dimensions read as:
\begin{eqnarray}
\lbrace{{G_{r+\eta}^+}{\bar{G}_{s-\eta}^-}}\rbrace&=&2L_{r+s}-(r-s+2\eta)J_{r+s}+(c/3)(({r+\eta})^2-1/4)\delta _{r+s,0}\nonumber\\
\lbrack L_n,L_m\rbrack &=&(n-m)L_{m+n}+{c\over {12}}m(m^2-1)\delta _{m+n,0}\nonumber\\
\lbrack L_n,G_{r\pm \eta}^{\pm} \rbrack &=&({n\over {2}} -{r\mp \eta})G_{n+{r\pm \eta}}^{\pm}\nonumber\\
\lbrack L_n,J_m \rbrack &=&-mJ_{m+n}\\
\lbrack J_m,J_n \rbrack &=&{c\over {3}}m \delta _{m+n,0}\nonumber\\
\lbrack J_n,G_{r+\eta}^{\pm} \rbrack &=&\pm G_{n+{r+\eta}}^{\pm}\nonumber\\
\lbrace{{G_{r+\eta}^+}{\bar{G}_{s-\eta}^+}}\rbrace&=&0\nonumber\\
\lbrace{{G_{r+\eta}^-}{\bar{G}_{s-\eta}^-}}\rbrace&=&0\nonumber
\end{eqnarray}
Similar eqs may be written down for the $N=4$ case. Eqs(5.3)
define a continuous one family parameter superconformal algebra to which we shall refer herebelow to as the $\eta$ sector and denote it as $[(1-2\eta )NS, 2\eta R]$. For $\eta=0$, one discovers the NS algebra and for $\eta ={1\over 2}$ one gets the R algebra. For $\eta$ ranging between zero and $1\over2$, one has the twisted sector. The $[(1-2\eta)NS, 2\eta R]$ twisted conformal algebra plays a crucial role in topological field theories [28,30 ]and allows to make spectacular transformations such as modifying the spins of the WS field operators by making appropriate choices of $\eta$. Taking the spectral parameter $ \eta ={1\over 2}$, a fermion transforms into a boson (scalar or vector) while taking $\eta ={1\over k}$, $k>2$, it becomes a WS parafermion of spin ($1\pm\eta$) depending on the $U(1)$ charge of the initial fermion. Putting back $\eta ={1\over k}$ into eqs(5.3), one gets amongst others: 
\begin{equation}
 2P_{-1}={\lbrace} G_{-{(k-1)/k}}^{+}, G_{-{1/k}}^{-} {\rbrace}+ {\lbrace}{G_{-{1/k}}^{+},G_{-{(k-1)/k}}^{-}}{\rbrace},
\end{equation}
together with:
\begin{equation}
\begin{array}{lcr}
0= {\lbrace} G_{-{1/k}}^{\pm},G_{-{1/k}}^{\pm}{\rbrace}\\
0={\lbrace} G_{-{(k-1)/k}}^{\pm}, G_{-{(k-1)/k}}^{\pm}{\rbrace}.
\end{array}
\end{equation}
Now comparing these relations with eqs(2.15) we obtained by thermal deformation of the $Z_k$ parafermoinic invariance, one discovers that they are quite similar. Eq (5.4) gives just a linearisation form of FSS which coincides with eqs(2.15) by setting $ k=3 $. Moreover eqs(5.5) show that $ G_{-{1/k}}^{\pm}$ are anticommuting operators in agreement with the result of [31]. Furthermore starting from eqs(5.4-5) and following the reasoning of section 2 which lead to the derivation of eqs(2.15), one sees that it is possible to reinterpret the minus charge carried by $G_{{(1-k)\over k}}^-$ as a $Z_k$ charge. So $G_{{(1-k)\over k}}^-$  may be viewed as as composite operator given by the product of $(k-1)$ $G_{-{1\over k}}^+$. This property is also supported by the fact that the N=2  superconformal currents have  mode expansion operators with twisted values. 
\begin{equation}
G^{\pm}(z_{1})\Phi_{m}(z_{2})=\sum z_{12}^{n-1\mp{p/k}}G _{-n\pm(p\pm 1)/k}^{\pm} \Phi_{m}(z_{2}).
\end{equation}
Using these modes operators, one may write for $k=3$ the following relations
\begin{equation}
G_{-{2\over3}}^-= G_{-{2\over3}}^+ {G_{0}^+}.
\end{equation}
Spectral flow of $N=2$  superconformal theories gives then an other way to build FSS algebras. In this regards, it is interesting to note that this spectral flow analysis maight also be used to rederive the so called FSUSY algebras considered recently in [29]. We suspect that the fractional quantum numbers considered in [29] when deriving FSUSY from special Verma modules of finite dimensional Lie algebras g could be rederived by taking fractional values of the spectral parameters $\eta$ of the corresponding Kac-Moody algebra $\hat g$. Recall in passing that under the spectral flow, the step generators $ J_{n}^\alpha$ and the Cartan ones $H_{n}^i$ of $\hat g$ transform as:
\begin{equation}
\begin{array}{lcr}
J_{n}^{\alpha} \to J_{n+\eta v.{\alpha}}^{\alpha}\\
H_{n}^i \to H_{n}^i + k\eta v^{i}\delta_{n,0},
  \end{array}  
\end{equation}
where ${\alpha} $ are the roots of $\hat g$ and v is a weight vector. This transformation shifts the eigenvalues of the $H_{n}^i$'s Cartan charge operators of $\hat g$. By an appropriate choice of the free parameters in the shifted weight ${{2k\eta}\over {\alpha^2}}\alpha^{i} v^{i}$ of ${2\over {\alpha^2}} \alpha^{i} H_0^{i}$, one recovers the   fractionality property of the quantum numbers used in the construction of FSUSY algebras[29] . This issue will be exhibited in more details in future occasion[32]. Now we turn to our main topic.

\section{Space-time invariance}

To analyse the space-time infinite dimensional symmetries on the boundary of $AdS_3$, one may follow the same strategy that we have used for the study of WS invariances. First identify the space time affine Kac-Moody symmetries and then consider the space time conformal invariance and eventually the Casimirs of higher ranks. In this section we shall simplify a little bit the analysis of space-time invariance and focus our attention on the conformal symmetry on $\partial ({AdS_3})$. Some specific features on space time Kac-Moody symmetries will also be given in due time.\\
We begin by noting that space time infinite invariances on the boundary of $AdS_3$ are intimately linked to the WS ones. For the case of a superstring propagating on $AdS_3{\times}S^3{\times}T^4$, we have already shown that there are various kinds of WS symmetries coming from the propagation on $AdS_3 , S^3$ and $T^4$ respectively. In the $\phi$ infinite limit,we want to show that one may use these WS symmetries to build new space time ones.\\ 
{\bf A}. {Conformal invariance}\\

First of all, note that the global part of the WS $SO(1,2){\times}{\bar{SO(1,2)}}$ affine invariance of a bosonic string on $AdS_3$, generated by $J_{0}^q$ and ${\bar{J}}_{0}^q; q=0,\pm 1$  may be  realized in different ways. A tricky way, which turns out to be crucial in building space-time conformal invariance, is given by the Wakimoto realization [22]. Classically, this representation reads in terms of the local coordinates $(\Phi, \gamma, \bar{\gamma})$ as follows:

\begin{equation}
\begin{array}{lcr}
J_{0}^0=\gamma {{\partial}/{\partial{\gamma}}}-
1/2{{\partial}/{\partial{\gamma}}},\\ 
J_{0}^-={\partial}/{\partial{\gamma}},\\
J_{0}^+={\gamma}^{2} {{\partial}/{\partial{\gamma}}}-{\gamma}
{{\partial}/{\partial{\Phi}}}-e^{-2{\Phi}}{\partial}/{\partial{\gamma}}.
\end{array}
\end{equation}
Similar relations are also valid for $\bar{J}_{0}^q$; they are obtained by substituting $\gamma$ by $\bar{\gamma}$. Quantum mechanically, the charge operators $J_{0}^q$ and $\bar{J}_{0}^q$ are given in terms of the Laurent mode operators of the quantum fields $\Phi, \gamma, \bar{\gamma}, \beta$ and $\bar{\beta}$ by using eqs(4.10) and performing the Cauchy integrations:
\begin{equation}
\begin{array}{lcr}
J_{0}^q=\int {dz\over 2i\pi}J^q(z)\\
\bar{J}_{0}^q=\int {d{\bar{z}} \over {2i\pi}}{\bar{J}}^{q}(z).
\end{array}
\end{equation}
To build the space time conformal invariance on the $AdS_3$ boundary, we proceed by steps. First suppose that there exists really a conformal symmetry on the boundary of $AdS_3$ and denote the space time Virasoro generators by $L_{n}$ and  $\bar{L}_{n}, n{\in} Z$. The ${L}_{n}$ and  ${\bar{L}}_{n}$, which should not be confused with the WS conformal mode generators, satisfy obviously the left and right Virasoro algebras.
\begin{equation}
\begin {array}{lcr}
[L_{n}, L_{m}]={(n-m)}L_{n+m}+{c/12}n{(n^{2}-1)}
{\delta _{n+m}}\\

[{\bar{L}}_{n},{\bar{L}}_{m}]=(n-m)\bar{L}_{n+m}+{{\bar{c}}/12}n(n^{2}-1)
{\delta_{n+m}}\\

[L_{n},{\bar{L}}_{m}]=0 .
\end {array}
\end{equation} 
The second step is to solve these eqs by using the string WS fields $(\Phi, \gamma, \bar{\gamma})$  on $AdS_3$. To do so, it is convenient to divide the above eqs into two blocks. The first block corresponds to set $ n=0,\pm 1$ in the generators $L_n$ and $\bar L_n$ of eqs(6.3). It describes the anomaly free projective subsymmetry the Virasoro algebra. The second block concerns the generators associated with the remaining values of n.\\
On the boundary of $ AdS_3$ obtained by taking the infinite limit of the $\Phi$  field, one solves the projective subsymmetry by natural identification of $L_{q}$ and ${\bar{L}}_q$ ; $q=0,\pm 1$ with the zero modes of the WS $so(1,2){\times}{\bar{so}}(1,2)$ affine invariance. In other words we have:
\begin{equation}
\begin{array}{lcr}
{L}_q&=&-\int {dz\over 2i\pi}J^q(z) =-J_{0}^q;\quad q=0,\pm 1\\
{\bar{L}}_q&=&\int {d{\bar{z}}\over 2i\pi}\bar{J}^q(z)
=-{\bar{J}}_{0}^q; \quad  q=0,\pm 1.
\end{array}
\end{equation}
Note that on the $AdS_3$ boundary, viewed as a complex plane parametrized by $( \gamma, \bar{\gamma} )$, the charge operators $J_{0}^-$ ( $L _{-1}$ ) and $\bar{J}_{0}^{-}(\bar{L}_{-1})$ taken in the Wakimoto representation  coincide respectively with the translation operators $P_{-}$ and $\bar{P}_{+}$:

\begin{equation}
\begin{array}{lcr}
P _{-}=L_{-}= {{\partial}/{\partial {\gamma}}}\\
P _{+}={\bar{L}}_{-}= {{\partial}/{\partial {\bar{\gamma}}}}.
\end{array}
\end{equation}
Eqs(6.4-5)) are interesting; they establish a link between the $L_{-}$ and ${\bar{L}}_{-}$ constants of motion of the boundary conformal field theory on $AdS_{3} $ on one hand and the $P _{-}(= P)$ and the $P _{+} (= \bar{P})$ translation generators of the ST extension of the $so(1,2)$ algebra on the other hand. We will turn to these relations  in the discussion of section 7.\\ 
To get the rigourous solution of the remaining Virasoro charge operators ${L}_{n}$ and  ${\bar{L}}_{n}$, one has to work hard. This is a lengthy and technical calculation which has been done in [10] in connection with the study of the $D_{1}/D_{5}$ brane system. Later on we shall give some indications on this method; for the time being we shall use an economic path to work out the solution. This is a less rigourous but tricky way to get the same result. This method is based on trying to extend the $L_{n}$ and $ \bar{L}_{n};\quad  n=0,{\pm 1}$ projective solution to arbitrary integers n using properties of the string WS fields near the boundary, dimensional arguments and similarities with the photon vertex operator in three dimensions. Indeed using the holomorphic property of $\gamma $ and $\bar{\gamma}$ eqs(4.9 ) as well as the space time dimensional arguments; 
\begin{equation}
\begin{array}{lcr}
 \lbrack \gamma \rbrack = -1;\quad  J_{sl(2)}^{0} = 0\\
 J_{sl(2)}^{-}= 1 ;\quad  J_{sl(2)}^{+}= -1, 
\end{array}
\end{equation}
it is not difficult to check that the following $L_{n}(\bar{L} _{n})$ expressions are good condidates:

\begin{equation}
{\bf L}_{n}=\int {dz\over {2i\pi}}\lbrack a_{0}{{\gamma}^{n}}J_{sl(2)}^{0}- { a_{-}\over 2} {{\gamma}^{n+1}}J_{sl(2)}^{-}+ { a_{+}\over 2} {{\gamma}^{n-1}}J_{sl(2)}^{+}\rbrack,
\end{equation}
and a similar relation for $\bar{L} _n$. To get the $a_i$ coefficients, one needs to impose constraints which may be obtained by using results of BRST analysis in QED in three dimensions. Following [9],the right constraints one has to impose on the $ a_i $' s are:

\begin{equation}
\begin{array}{lcr}
    n a_{0}  + (n+1 ) a_{-} + (n-1 ) a_{+}= 0\\
    J^{0}{\gamma}-(1/2) J^{-}{\gamma^2}-(1/2) J^{+} = 0.
\end{array}
\end{equation}
The solution of the first constraint of these eqs reproducing the projective generators (6.4) is as follows:

\begin{equation}
\begin{array}{lcr}
 a_{0}= ({n^2}-1)\\
a_{-}= n(n-1)\\
a_{+}= n(n+1)
\end {array}
\end{equation}
Moreover using the second constraint of eqs(6.8) to express $J_{sl(2)}^{+}(z)$ in terms of $J_{sl(2)}^{0}(z)$ and $J_{sl(2)}^{-}(z)$; then putting back into eqs(6.7), we find: 

\begin{equation}
{\bf L}_{n}=\int {dz\over {2i\pi}}\lbrack {-( n+1){{\gamma}^{n}}J_{sl(2)}^{0}+ n {{\gamma}^{n+1}}J_{sl(2)}^{-}}\rbrack .
\end{equation}
Eqs (6.4) and (6.10) define the space time Virasoro algebra on the boundary of $AdS_3$.
{\bf B}. {comments}\\ 
Having built the $L_n$'s space time Virasoro generators, one may be interested in determining the space-time energy momentum tensors $T({\gamma})$ and $\bar{T}{({\bar\gamma})}$ of the BCFT on $AdS_3$. It turns out that the right form of the space-time energy momentum tensor depends moreover on  auxiliary complex variables $(y,\bar{y})$ so that the space time energy momentum tensor has now two arguments; i.e: $T=T(y,\gamma)$ and $\bar{T}= \bar{T}({\bar{y}},\bar{\gamma})$. Following [10], $T(y,\gamma )$ and $\bar{T}({\bar{y}},\bar{\gamma})$ read as:

\begin{equation}
\begin{array}{lcr}
T(y,\gamma)= \int {dz\over{2i\pi}}\lbrack {{\partial_{y}{J(y,\gamma)}}\over{{(y-\gamma)}^{2}}}
-{{{{\partial^{2}}_{y}}J(y,\gamma)}\over{(y-{\gamma})}}\rbrack\\
\bar{T}({\bar{y}},{\bar\gamma})=\int {d{\bar{z}}\over{2i\pi}}\lbrack {\partial_{\bar{y}{J(\bar{y},\bar{\gamma})}}\over{(\bar{y}-\bar{\gamma})}^{2}}-{{{\partial^{2}}_{\bar{y}}{J(\bar{y},\bar{\gamma})}}\over{(\bar{y}-\bar{\gamma})}} \rbrack,
\end {array}
\end{equation}
where the currents $J(y,\gamma)$ and $J(\bar{y},\bar\gamma)$ are given by:
\begin{equation}
\begin {array}{lcr}
J(y,\gamma )=-J^{+}(y,\gamma)=2yJ^{0}(\gamma)-J^{+}(\gamma)-y^{2}J^{-}(\gamma).
\end{array}
\end{equation}
In connection to these eqs, it is interesting to note that the conserved currents $J^{q}(y,\gamma)$ and $J^{q}(\bar{y},\bar{\gamma})$ are related to the WS affine Kac-Moody ones on $AdS_3$ as follows:

\begin{equation}
\begin{array}{lcr}
J^{+}(y,\gamma)= e^{-yJ_{0}^{-}}J^{+}(\gamma)e^{yJ_{0}^{-}}=J^{+}(\gamma)-2yJ^{0}(\gamma)+y^{2}J^{-}(\gamma)\\
J^{0}(y,\gamma)=e^{-yJ_{0}^{-}}J^{0}(\gamma)e^{yJ_{0}^{-}}=J^{0}(\gamma)-yJ^{-}(\gamma)={-{1\over 2}}{\partial _z}J^{+}(y,\gamma)\\
J^{-}(y,\gamma)=e^{-yJ_{0}^{-}}J^{-}(\gamma)e^{yJ_{0}^{-}}=J^{-}(\gamma)={1\over 2}{{\partial }^{2} _z}J^{+}(y,\gamma).
\end {array}
\end{equation}
and analogous eqs for $J^{q}(\bar{y},\bar{\gamma})$. Putting eqs(6.12) back into eqs(6.11) and expanding in power series of $\gamma \over{y}$, one discovers the $ L_n$ space time Virasoro generators given by eqs( 6.10).\\
The second comment we want to make concerns the building of space time affine Kac-Moody symmeties out of the WS ones. Staring from WS conserved currents $E_{ws}^a(z)$, which may be thought of as $ J_{sl(2)}^{q}(z)$, and going to the boundary of $AdS_3$, the corresponding space time affine Kac-Moody currents $E_{space time}^a (y,\gamma)$ read as:

\begin{equation}
E_{space time}^a(y,\gamma)= \oint  {dz\over 2i\pi}\lbrack {{E_{ws}^{a}(z)}\over{(y-\gamma(z))}}\rbrack.
\end{equation}
Expanding this eq in powers of $y\over {\gamma}$ or $\gamma \over {y}$, one gets the space time affine Kac-Moody modes:

\begin{equation}
E_{n}^{a,space time}= \oint {dz\over {2i\pi}}\lbrack E_{ws}^{a}(z){\gamma}^{n}\rbrack.
\end{equation}
The third comment we want to make concerns superstrings on $AdS_{3}\times {S^3}\times{T^4}$. In addition to the bosonic sector,there are moreover contributions coming from the WS fermions $\psi_{ws}(z)$. On the $AdS_3$ space for which the WS fermions $\psi _{ws}^{q}(z)$, $q=o,\pm$ transform in the $SO(1,2)$ adjoint, the total level k $SO(1,2)$ currents $J_{sl(2),Total}^{q}(z)$ now have two contributions: a level (k+2) bosonic current $J_{sl(2),Bose}^{q}(z)$ and a level (-2) fermionic current $J_{sl(2),Fermi}^{q}(z)$. The same construction may also be done for both $S^3$ and $T^4$.
Note finally that in the limit $\phi$ goes to infinity, the space time conformal symmetry of a superstring propagating on $AdS_{3}\times {S^{3}}\times {T^{4}}$ form a $N=4$ conformal invariance.
\section{Discussions and Conclusion}
We have learned hereabove that on $AdS_{3}\times {N^d}$ may live various boundary conformal field theories depending on the choice of the d-dimensional compact manifold $N^d$. In the case of critical models of (super) strings propagating on $AdS_{3}\times {N^d}$, we have studied two examples: (i) $N^d$ is given by  ${T^23}$ torus. (ii) $N^d$ is given by ${S^3}\times{T^4}$. The first example describes a bosonic BCFT while the second one describes a $ N=4$ BCFT. One may also considers other choices of $ N^d$ and build other BCFT's.\\ If one forgets about string dynamics as well as the nature of the compact manifold $ N$ and just retains that on $\partial({AdS_{3}})$ lives a conformal structure, one may consider its highest weight representations which read in general as: 

\begin{equation}
\begin{array}{lcr}
{L _{0}}\vert {h,{\bar{h}}}\rangle=h {\vert {h,{\bar{h}}}\rangle},\\
{L_n}{\vert {h,{\bar{h}}}\rangle}=0;\quad n\geq 1\\
{\bar L_{0}}{\vert {h,{\bar{h}}}\rangle}=\bar{h}{\vert {h,{\bar{h}}}\rangle},\\
{{\bar{L}}_{n}}{\vert {h,{\bar{h}}}\rangle}=0;\quad n{\geq}1,\\
cI\vert {h,{\bar{h}}}\rangle=c\vert {h,{\bar{h}}}\rangle,
\end{array}
\end{equation}
where $\vert {h,{\bar{h}}}\rangle$ are Virasoro primary states. A priori the central charge c and the conformal weights h and $\bar{h}$ of these representations are arbitrary. However requiring unitary conditions, the parameters c, h and $\bar{h}$ are subject to constraints which become more stronger if one imposes extra symmetries such as supersymmetry or parafermionic invariance. Having these details in mind, one may also build descendant states $\vert {{h+n},{\bar{h}+\bar{n}}}\rangle$ of $\vert {h,{\bar{h}}}\rangle$ from the primary ones as follows: 
\begin{equation}
 \vert {{h+n},{\bar{h}+\bar{n}}}\rangle={\sum_ {\stackrel{n=\sum {\alpha _{i}} n_{i}}{\bar{n}=\sum {\beta _{j}} n_{j}}}\lambda _{\{\alpha_{i}\} \{\beta _{j}\} }{(\Pi _{i}L_{-n_{i}}^{\alpha {i}})(\Pi _{j}\bar{L}_{-n_{j}}^{\beta {j}}})} \vert {h,{\bar{h}}}\rangle.
\end{equation}
where the $\alpha_{i}$'s and $\beta_{j}$'s are positive integers and $\lambda _{\alpha \beta }$ are C-numbers which we use to denote the collective coefficients $\lambda _{\{\alpha_{i}\} \{\beta _{j}\} } $ of the decomposition eq(7.2).
They satisfy the following obvious relations.

\begin{equation}
\begin {array}{lcr}
{L _0}{\vert {{h+n},{\bar{h}+\bar{n}}}\rangle}=(h+n){\vert {{h+n},{\bar{h}+\bar{n}}}\rangle}\\

{L_{\pm}}{\vert {{h+n},{\bar{h}+\bar{n}}}\rangle}=a_{\pm}(h,n){\vert {{h\pm n},{\bar{h}\pm \bar{n}}}\rangle}\\

{{\bar{L}} _0}{\vert {{h+n},{\bar{h}+\bar{n}}}\rangle}=({\bar{h}}+{\bar{n}}){\vert {{h+n},{\bar{h}+\bar{n}}}\rangle}\\

{{\bar{L}}_{\pm}}{\vert {{h+n},{\bar{h}+\bar{n}}}\rangle}
={{\bar{a}}_{\pm}}(\bar{h},\bar{n}){\vert {h\pm n},{\bar{h}\pm \bar{n}}\rangle}
\end {array}
\end{equation}
where $a_{\pm}(h,n)$ and $\bar{a} _{\pm}(\bar{h},{\bar{n}})$  are normalization factors. Making an appropriate choice of the $\lambda _{\alpha \beta }$ coefficients and taking the $a_{\pm}(h,n)$ and $\bar{a} _{\pm}(h,\bar{n})$ coefficients as given herebelow,   
\begin{equation}
\begin {array}{lcr}
a_{-}(h,n)=\sqrt{(2h+n)(n+1)}\\
a_{+}(h,n)=\sqrt{(2h+n-1)n},
\end {array}
\end{equation}
one can get the two $so(1,2)$ modules used in building RdTS supersymmetry.
Note that the descendant states $\vert {{h+n},{\bar{h}+\bar{n}}}\rangle$ are also eigenstates of the spin $({L _0}-{{\bar{L}}_0})$ and conformal scale $(L_{0}+ \bar{L}_{0})$  operators of eigenvalues $s=[(h- \bar{h})+ (n-{\bar{n}})]$ and $\Delta=[(h+{\bar{h}})+ (n+{\bar{n}})]$ respectively.\\
We conclude this study by saying that the RdTS extension of Poincar\'e invariance in (1+2) dimensions that we have been describing is a special kind of FSS algebra. Like for FSS invariances, the RdTS generalised algebra may be also viewed as a residual symmetry of a boundary conformal invariance living on (1+2) space time manifolds. The RdTS supersymmetry we have described is a special FSS because it is related to a deformation of the space time boundary conformal invariance on $AdS_3$. In the end of this study, we should say that the explicit analysis of this paper has been plausible due to the particular properties of the $AdS_3$ geometry: (a) the $AdS_3$ manifold carries naturally a $so(1,2)$ affine invariance which has various realisation ways. (b) the Wakimoto realisation of the $SO(1,2)$ affine symmetry which on one hand relates its zero mode to the projective symmetry of a BCFT on $AdS_3$ and on the other hand links the $L_{-}$ and $\bar L_{-}$ to the translation operators on $\partial {AdS_3}$. (c)the correspondance between WS and space time symmetries which plays a crucial role in analysing the various kinds of symmetries living on $\partial {AdS_3}$. \\
At last we would like to note that this study maight find a natural application in FQH systems formulated as an effective Chern-Simon gauge theory. In this model, the physics in the bulk is roughly speaking described by a $ (1+2)$ dimensional $ U(1)^n$ gauge while the theory the edge excitations of FQH liquids are described by a boundary conformal field theory. We plan to extend the results of this paper to the case of FQH droplets in a future occasion.\\
\bigskip
\section{Aknowledgements}
The authors would like to thank Dr Rausch de Traubenberg for discussions, suggestions and for reading the manuscript. EHS would like to thank E.M Sahraoui for earlier collaboration on strings on Anti-de Sitter space and A Belhaj for discussions.
 This research work is supported by the program PARS Physique 27 under contract 372-98 CNR. 

\newpage

{\bf {References}}
\begin{enumerate}

\item[[1]]M.Rausch de Traubenberg and M.J. Slupinski,
          Mod.Phys.Lett. A12 (1997) 3051-3066.
\item[[2]] D.Friedan, Z.Qiu, S.Shenker, Phys.Rev.Lett52(1984)1575. 
   \item[[3]] G. Mussardo and P. Simonetti, Int.Jour.Mod.Phys.A9(1994) 3307-3338. S.Cecotti, C. Vafa Commun. Math. Phys. 157 (1993) 139-178
\item[[4]] G.Mussardo Phys.Rep 182 (1993) 
\item[[5]] R.E.Prange and S.M. Girvin, The Quantum Hall effect (Springer, New York, 1987),R.B.Laughlin,Phys.rev.Lett. 50 (1983) 1395.
\item[[6]] X.G.Wen. Topological orders and Edge Excitations in FQH states. Cond-mat/9506066.
\item[[7]] X.G.Wen,A. Zee,Field Theory, Topology and Condensed Matter Physics,Proceedings of the Ninth Chris Engelbrecht Summer School in Theoretical Physics,Held at Storms River Mouth, Tsitsikamma National Park,South Africa, 17-28 January 1994 (Springer-Verlag, 1995, Hendrik B Geyer(Ed)).
\item[[8]] N.Seiberg, E.Witten The $D_1/D_5$ System and Singular CFT,hep-th/9903224.
\item[[9]]A.Giveon,D.Kutasov and N.Seiberg, Comments on String Theory on $AdS_3$, hep-th/9806194.        \item[[10]] D.Kutasov, N.Seiberg, More Comments on String Theory on $AdS_3$,hep-th/9903219 , JHEP 9904 (1999) 008
\item[[11]]A.Leclair,C.Vafa Nucl.Phys. B401 (1993) 413.
          D.Bernard, A Leclair,Nucl.Phys.B340(1990)712;
          Phys.Lett B247 (1991)309; 
          Commun. Math. Phys.142 99.
\item[[12]]E.H. Saidi, M.B. Sedra and J. Zerouaoui,Class.Quant.                       Grav.12(1995)1567-1580.
\item[[13]]A.Perez,M Rausch de Traubenberg,P.Simon, Nucl.Phys.B 482 (1996)352
\item[[14]]H. Ahmedov, O.F. Dayi,
 Non-Abelian Fractional Supersymmetry in Two Dimensions,math.QA/9905164. 
 Omer F. Dayi, $U_q(sl(2)$ as Dynamical Symmetry Algebra of the Quantum Hall Effect, math.QA/9803032
\item[[15]]A. Jellal, Mod. Phys. Lett. A14 (1999) 2253,
M.Rachidi, E.H.Saidi, J.Zerouaoui Phys.Lett.B409 (1997) 349-354.
\item[[16]]I.Benkaddour,E.H. Saidi Class.Quantum. Grav.16 (1999)1793-1804.
\item[[17]]A.Elfallah, E.H Saidi, J.Zerouaoui Phys.Lett.B468 (1999)86-95.
\item[[18]]A.M Zamolodchikov,V.A Fateev(1985) Sov.Phy-JETP 62 215 .
\item[[19]]H.Chakir,A.Elfallah, E.H.Saidi, Mod.Phys.LettA38(1995)2931.
\item[[20]]P.C.Argyres, S.H-H.Tye, Commun.Math.Phys. 159(1993),471.
           P.C.Argyres,A.Leclair, S.H-H.Tye, Phys. Lett.B253(1991).
           P.C.Argyres, S.H-H.Tye, Phys.Rev. Lett.67(1991),3339.
\item[[21]]H.Chakir,A.Elfallah,E.H.Saidi,Class.Quant.Grav.14(1997)2049.
\item[[22]]M.Wakimoto, Comm.Math.Phys.104(1986)605.
\item[[23]] M.Rausch de Traubenberg. Lectures delivred at the workshop on non commutavive Geometry and Superstring theory, Rabat 16-17 June(2000). 
\item[[24]]J.Maldacena, Adv.Theor.Math.Phys.2 (1997)231.hep-th/9711200.
\item[[25]]E.Witten, Heterotic String Conformal Field Theory And A-D-E Singularities,hep-th/9909229.

\item[[26]]J.Balog,L.O'Raifeartaigh,P. Forgacs,A.Wipf, Nucl.Phys.B325 (1989)225.
\item[[27]]C.Vafa, N.P.Warner, Phys.Lett. B218(1989)51.
W. Lerche, C.Vafa, N.P.Warner Nucl.Phys. B324 (1989)427. 
\item[[28]]E.Witten, Introduction to cohomological field theories, lectures given at Conf. on Topological Methods in Quantum Field Theory (Trieste, June 1990), Int. J.Mod. Phys. A6(1991)2775. J. Sonnenschein, Topological Quantum Field Theories, moduli spaces, and flat connections, Phys. Rev. D 42 (1990) 2080.M.Bershadsky, S.Cecotti, H.Ooguri, C.Vafa 
 Nucl.Phys. B405 (1993) 279-304.
  T. Eguchi, Y. Yamada and S.-K. Yang, Topological Field Theories and the Period Integrals, hep-th/9304121 : Mod. Phys. Lett. A8 (1993) 1627-1638
\item[[29]]M.Rausch de Traubenberg, M. J. Slupinski
Fractional Supersymmetry and Fth-Roots of Representations,J.Math.Phys.41(2000)4556-4571.
\item[[30]]S.Gukov,C.Vafa, E.Witten;  
CFT's From Calabi-Yau Four-folds,hep-th/9906070.
\item[[31]]I.Benkaddour, E.H. Saidi, Class. Quantum Grav.16(1999)1793-1804.
\item[[32]] Work in progress.
\end{enumerate}

\end{document}